\documentclass[english,10pt,aps,pra,showpacs,twocolumn,floatfix]{revtex4-1}

\usepackage[T1]{fontenc}
\usepackage{amsmath}
\usepackage{amssymb}
\usepackage{mathtools}
\usepackage{amsfonts}
\usepackage{bbm}
\usepackage{epsfig}
\usepackage{grffile}
\usepackage{babel}
\usepackage{graphics}

\usepackage[usenames,dvipsnames]{xcolor}
\definecolor{grey}{rgb}{.5,.5,.5}
\definecolor{dblue}{rgb}{0,0,.5}
\definecolor{dgreen}{rgb}{0,.65,0}

\usepackage[colorlinks=true,citecolor=dblue,linkcolor=dblue,urlcolor=dblue]{hyperref}
\usepackage[all]{hypcap}
\newcommand{\Footnote}[1]{\footnote{\unexpanded{#1}}}

\newcommand{\Tr}{\operatorname{Tr}}
\newcommand{\tot}{{\mathrm{tot}}}
\newcommand{\bra}{\langle}
\newcommand{\ket}{\rangle}
\newcommand{\mc}[1]{\mathcal{#1}}
\newcommand{\ce}{{\mathrm{c}}}
\newcommand{\gce}{{\mathrm{gc}}}

\renewcommand{\H}{\mc{H}}
\newcommand{\hH}{\hat{H}}
\newcommand{\hK}{\hat{K}}
\newcommand{\hS}{\hat{S}}
\newcommand{\hq}{\hat{q}}
\newcommand{\hQ}{\hat{Q}}
\newcommand{\hO}{\hat{O}}
\newcommand{\dm}{{\hat{\varrho}}}
\renewcommand{\vec}[1]{{\boldsymbol{#1}}}

\newcommand{\vn}{{\vec{n}}}
\newcommand{\vs}{{\vec{\sigma}}}
\newcommand{\ua}{{\uparrow}}
\newcommand{\da}{{\downarrow}}
\newcommand{\SWAP}{\text{SWAP}}

\newcommand{\duke} {Department of Physics, Duke University, Durham, North Carolina 27708, USA}

\begin{document}

\newcommand{\titlefont}{\fontfamily{ptm}\selectfont}
\title{\titlefont Symmetric minimally entangled typical thermal states for canonical and grand-canonical ensembles}
\author{Moritz Binder}
\affiliation{\duke}
\author{Thomas Barthel}
\affiliation{\duke}
\date{January 3, 2017}

\begin{abstract}
Based on the density matrix renormalization group (DMRG), strongly correlated quantum many-body systems at finite temperatures can be simulated by sampling over a certain class of pure matrix product states (MPS) called minimally entangled typical thermal states (METTS). When a system features symmetries, these can be utilized to substantially reduce MPS computation costs. It is conceptually straightforward to simulate canonical ensembles using symmetric METTS. In practice, it is important to alternate between different symmetric collapse bases to decrease autocorrelations in the Markov chain of METTS. To this purpose, we introduce symmetric Fourier and Haar-random block bases that are \emph{efficiently mixing}. We also show how grand-canonical ensembles can be simulated efficiently with symmetric METTS. We demonstrate these approaches for spin-$1/2$ XXZ chains and discuss how the choice of the collapse bases influences autocorrelations as well as the distribution of measurement values and, hence, convergence speeds.
\end{abstract}


\maketitle

\section{Introduction}
The density matrix renormalization group (DMRG) is a powerful numerical technique for the simulation of one-dimensional (1D) strongly correlated quantum systems \cite{White1992-69,White1993-48,Schollwoeck2005}. Its concise formulation in terms of matrix product states (MPS) \cite{Fannes1992-144,Rommer1997-55,Schollwoeck2011-326} provides a framework for the efficient computation of ground states and the real-time evolution of pure states. Three quite different approaches were developed that extend it to the simulation of finite-temperature properties. The historically first was the quantum transfer-matrix renormalization group \cite{Nishino1995-64,Bursill1996-8,Shibata1997-66,Wang1997-56} which has some technical complications such as the non-Hermiticity of the quantum transfer matrix.
The second one relies on a purification of the mixed state \cite{Uhlmann1976,Uhlmann1986,Nielsen2000} that can be encoded in matrix product form \cite{Verstraete2004-93,Barthel2016-94}. These matrix product purifications (MPP) were successfully applied to calculate for example finite-temperature correlation and response functions of quantum spin chains \cite{Feiguin2005-72,Barthel2005,Barthel2009-79b}. As, e.g., described in Ref.~\cite{Barthel2013-15}, the employed MPPs are in one-to-one relation with matrix product density operators (MPDO) \cite{Zwolak2004-93}.
However, MPP computations remain challenging, because the simulation on an enlarged Hilbert space can lead to a considerable growth of entanglement, making them costly at low temperatures. More recently, an alternative approach was developed that avoids the purification and hence the enlarged Hilbert space. Instead, one samples over a cleverly chosen set of pure quantum states, called minimally entangled typical thermal states (METTS) \cite{White2009-102,Stoudenmire2010-12}. The efficiency of the METTS algorithm is limited by the statistical error induced by the sampling \cite{Binder2015-92}.

In this paper, we describe and demonstrate how symmetries can be utilized to improve the efficiency of the METTS algorithm. Recently, it was shown in Ref.~\cite{Bruognolo2015-92} how grand-canonical METTS simulations of response functions can be made substantially more efficient by switching to symmetric states just before the real-time evolution. However, the actual METTS sampling was unmodified, i.e., symmetries were not employed in the imaginary-time evolution and transitions.

Here, we discuss symmetric METTS algorithms for both canonical and grand-canonical ensembles. If the system and its environment exchange energy and there is a conserved quantity $\hQ$, the equilibrium state of the system is given by the (here, unnormalized) \emph{canonical ensemble}
\begin{equation}\label{eq:CE}
	\dm^\ce_{\beta,Q}:= e^{-\beta\hH_Q} \quad\text{on $\H_Q$}
\end{equation}
with $\hH_Q$ being the projection of the Hamiltonian onto the quantum number $Q$ subspace $\H_Q$ of the full Hilbert space $\H=\bigoplus_Q \H_Q$.
Similarly, if system and environment also exchange the quantity associated with $\hQ$, the equilibrium state is the \emph{grand-canonical ensemble}
\begin{equation}\label{eq:GCE}
	\dm^\gce_{\beta,\alpha}= e^{-\beta(\hH+\alpha\hQ)}.
\end{equation}
Here, the Lagrange multiplier $\alpha$ fixes the expectation value of $\hQ$. In more complex cases with multiple conserved quantities $\hQ^{(j)}$, one can also consider ensembles like $\exp[{-\beta(\hH_{Q^{(1)}}+\alpha_2\hQ^{(2)})}]$.

The transitions in the Markov chain of METTS samples are determined by projective measurements with respect to a collapse basis that can be freely chosen. This choice strongly affects the statistical properties of the resulting samples \cite{Stoudenmire2010-12,Binder2015-92}. In this work, we introduce novel collapse bases for symmetric METTS simulations. To be able to conserve global quantum numbers $Q$ and to increase efficiency, we go beyond bases of single-site product states and carry out the projective measurements on blocks of several sites. We discuss the influence of the collapse basis choice on the convergence of the algorithm and introduce Fourier and Haar-random block bases which are, as we call it, \emph{efficiently mixing}.

The structure of this work is as follows. In Sec.~\ref{sec:symmetries} we review how symmetries can be utilized in the matrix product state representation to achieve a significant speedup in the simulations. We briefly summarize the original METTS algorithm without the use of symmetries in Sec.~\ref{sec:METTS}. In Sec.~\ref{sec:CE}, we discuss how to use symmetries in the simulation of canonical ensembles. We go on to introduce maximally and efficiently mixing (symmetric) collapse bases in Sec.~\ref{sec:bases} and summarize the factors influencing convergence speeds in Sec.~\ref{sec:factors}. Section~\ref{sec:CE-XXZ} applies the techniques to spin-$1/2$ XXZ chains in the canonical ensemble. A symmetric METTS algorithm for the simulation of grand-canonical ensembles is introduced and demonstrated in Sec.~\ref{sec:GCE}. We summarize and conclude in Sec.~\ref{sec:conclusions}.

\section{Symmetries in matrix product states} \label{sec:symmetries}
Let us consider a lattice system with $L$ sites and orthonormal on-site basis states $\{|\sigma_i\ket\,|\,\sigma_i=1,\dotsc,d\}$. Matrix product states (MPS) for the system have the form
\begin{equation}\label{eq:MPS}
	|\psi\ket = \sum_\vs A^{\sigma_1}_1 A^{\sigma_2}_2\dotsb A^{\sigma_L}_L|\vs\ket,
\end{equation}
with $\vs:=(\sigma_1,\dotsc,\sigma_L)$ and $D_{i-1}\times D_{i}$ matrices $A^{\sigma_i}_i$. The $D_i$ are also called bond dimensions and we require $D_0=D_{L}=1$ such that the matrix product yields a scalar coefficient.

We are concerned with states $|\psi\ket$ that are eigenstates of a conserved quantity $\hQ$. For simplicity, we restrict our considerations to a single Abelian symmetry such as conservation of total particle number or magnetization. However, everything generalizes in a very similar manner to the cases of multiple conserved quantities and non-Abelian symmetries \cite{McCulloch2002-57,Weichselbaum2012-327}. For the latter, one exploits that dependencies inside each multiplet are given by Clebsch-Gordan coefficients as exemplified by the Wigner-Eckart theorem.

For an Abelian symmetry, the conserved quantity may have the form $\hQ = \sum_i \hq_i$ with $\hq_i|\sigma_i\ket = q(\sigma_i)|\sigma_i\ket$ \Footnote{The total quantum number can arise less trivially than by summing single-site quantum numbers, e.g., by addition modulo $N$ for a cyclic group of order $N$.}. We can construct an MPS \eqref{eq:MPS} with quantum number $Q$ by imposing selection rules on the tensor elements $[A^{\sigma_i}_i]_{a_i,b_i}$. Specifically, one assigns quantum numbers $q(a_i)$ and $q(b_i)$ to the matrix indices and imposes selection rules like
\begin{equation} \label{eq:selection_rule}
 [A^{\sigma_i}_i]_{a_i,b_i} \neq 0 \quad\text{only if}\quad q(a_i)+q(\sigma_i)=q(b_i).
\end{equation}
\begin{center}
\includegraphics[width=0.35\columnwidth]{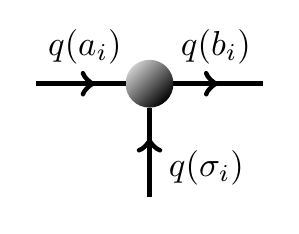}
\end{center}
Setting $q(a_1)=0$ and $q(b_L)=Q$, this ensures that $\hQ|\psi\ket=Q|\psi\ket$.

Explicitly enforcing these conditions by decomposing the tensors into symmetry blocks leads to a significant speedup and improved accuracy of the MPS algorithms. The numerically most costly operations are typically singular value decompositions (SVD) of the tensors in the MPS. These then reduce to cheaper SVDs of the symmetry blocks.

\section{Minimally entangled typical thermal states} \label{sec:METTS}
The strategy employed in the minimally entangled typical thermal states (METTS) algorithm \cite{White2009-102} is to decompose the thermal density matrix $\dm_\beta:=\exp(-\beta\hK)$, with $\hK=\hH_Q$ or $\hK=\hH-\alpha\hQ$, into a sum of projectors
\begin{equation*}\label{eq:METTSensemble}
	\dm_\beta = \sum_\vn P_\vn |\phi_\vn\ket\bra\phi_\vn|,
\end{equation*}
with METTS
\begin{equation*}
	|\phi_\vn\ket := \frac{1}{\sqrt{P_\vn}}e^{-\beta \hK/2}|\vn\ket,\quad
	P_\vn:= \bra \vn | e^{-\beta \hK} | \vn\ket.
\end{equation*}
Here, $\mc{B}:=\{|\vn\ket\}$ is an appropriate orthonormal basis for the full system with $\bra\vn|\vn'\ket=\delta_{\vn,\vn'}$. As discussed later, we choose \emph{block product states}. Block product states may be entangled within blocks of a certain number of lattice sites but not across the block boundaries. Correspondingly, the MPS bond dimensions $D_i$ are 1 at the block boundaries.

As described in Refs.~\cite{White2009-102,Stoudenmire2010-12,Binder2015-92}, one can efficiently generate a Markov chain
\begin{equation*}
	\phi_{\vn}\to\phi_{\vn'}\to\phi_{\vn''}\to\dots
\end{equation*}
of METTS in MPS form according to their (unnormalized) probabilities $P_\vn$ by repeated imaginary-time evolution steps $|\vn\ket\to |\phi_\vn\ket$ and projective measurements $|\phi_\vn\ket\to|\vn'\ket$. The evolution step can 
be executed with time-dependent DMRG (tDRMG) \cite{White2004,Daley2004}. The transition probabilities $p_{\vn \to \vn'} = |\bra\vn'|\phi_\vn\ket|^2$ obey detailed balance
\begin{equation}\label{eq:balance1}
	P_\vn p_{\vn\to\vn'} = |\bra\vn'|e^{-\beta\hK/2}|\vn\ket|^2 = P_{\vn'}p_{\vn'\to\vn}.
\end{equation}
Thermal expectation values $\bra \hat{O} \ket_\beta=\Tr(\dm_\beta\hat{O})/\Tr(\dm_\beta)$ can then be computed by averaging $\bra\phi_{\vn^\nu}| \hat{O}|\phi_{\vn^\nu}\ket$ over the Markov chain. If the states $|\vn\ket$ are (block) product states, the projective measurements can be executed in a sweep through the lattice by doing local projective measurements \cite{White2009-102,Stoudenmire2010-12,Binder2015-92}.

In the following, we discuss how symmetries can be utilized in METTS simulations for canonical and grand-canonical ensembles, i.e., how the conservation of $\hQ=\sum_i\hq_i$ eigenvalues can be used to substantially reduce computation costs.
Please note that Ref.~\cite{Bruognolo2015-92} shows how to produce symmetric METTS for the grand-canonical ensemble for the evaluation of time-dependent quantities (study of quenches or response functions). To this purpose, non-symmetric METTS, which are not $\hQ$ eigenstates, have been generated. Symmetric METTS are then obtained from these in subsequent symmetric collapses. While this does not provide any computational advantage for the imaginary-time evolution and the evaluation of static quantities, it can make subsequent real-time evolutions of the METTS, in which the symmetries are exploited, much more efficient \cite{Bruognolo2015-92}. What is described in the following offers an efficient way to already utilize symmetries during the imaginary-time evolution.

\section{Symmetries for the canonical ensemble} \label{sec:CE}
Conceptually, it is straightforward to simulate canonical ensembles \eqref{eq:CE} using symmetric METTS. One simply needs to restrict the initial state and the collapse basis $\{|\vn\ket\}$ to the correct symmetry sector. In particular, one should work with an orthonormal basis
\begin{equation}\label{eq:symmBasis}
	\{|\vn\ket\,\,|\,\,\hQ|\vn\ket=Q|\vn\ket\}
\end{equation}
of $\H_Q$. If these states are (block) product states, they can be easily encoded as symmetric MPS \eqref{eq:MPS} with small bond dimensions, where matrix elements obey the constraint \eqref{eq:selection_rule}. As the Hamiltonian $\hH_Q$ commutes with $\hQ$, the symmetry constraints on the MPS also hold during the imaginary-time evolution $|\vn\ket\to |\phi_{\vn}\ket$. Now, the projective measurements $|\phi_\vn\ket\to|\vn'\ket$ need to be done such that also $|\vn'\ket$ has quantum number $Q$. If we use a symmetric collapse basis (i.e., every basis state is a $\hQ$ eigenstate), we always stay in the same symmetry sector as transition probabilities $p_{\vn\to\vn'}=|\bra\vn'|\phi_\vn\ket|^2$, in the projective measurements, vanish for states $|\vn'\ket$ with quantum number $Q'\neq Q$. Inside the symmetry sector with quantum number $Q$, detailed balance is fulfilled as in Eq.~\eqref{eq:balance1}.

In practice, using the same symmetric collapse basis \eqref{eq:symmBasis} for every transition in the Markov chain can however be very inefficient, because it often leads to strong autocorrelations between subsequent METTS samples. This is obvious for infinite temperature, where we would be stuck in the initial state of the Markov chain. To arrive at an efficient algorithm, one needs to alternate between different symmetric bases. To this purpose, we introduce novel symmetric collapse bases and discuss their properties in the following (Sec.~\ref{sec:bases}).

After completion of this work, we noticed Ref.~\cite{Lacki2015-91}. To our knowledge it is the only previous work trying to simulate canonical ensembles with METTS. In particular, a canonical ensemble for the Bose-Hubbard model in the gapped Mott regime was simulated, using only the $\{\hat{n}_i\}$ eigenbasis, i.e., Fock states. Because of the strong autocorrelations, only every 200th METTS sample was included in the final ensemble.

\section{Efficient collapse bases} \label{sec:bases}
It is possible and often advantageous to switch between different collapse bases in order to decrease autocorrelation times in the Markov chain. For example, one can do projective measurements using a basis $\{|\vn\ket\}$ for all odd iteration steps and a second basis $\{|\tilde{\vn}\ket\}$ for all even iteration steps. Detailed balance is still fulfilled in every second iteration step, as
\begin{equation}\textstyle
	P_\vn \sum_{\tilde{\vn}}p_{\vn\to\tilde{\vn}}p_{\tilde{\vn}\to\vn'}
	=P_{\vn'}\sum_{\tilde{\vn}}p_{\vn'\to\tilde{\vn}}p_{\tilde{\vn}\to\vn},
\end{equation}
where $|\vn\ket$ and $|\vn'\ket$ are from basis 1 and $|\tilde{\vn}\ket$ from basis 2.

A simple example is to collapse alternatingly to $\{\hS^z_i\}$ and $\{\hS^x_i\}$ eigenstates, respectively, for a spin-$1/2$ system as described in Ref.~\cite{Stoudenmire2010-12}. For a general system with a $d$-dimensional local state space (e.g., the Bose-Hubbard model with a maximum of $n_{\max}=d-1$ particles per site), one can generate Haar-random collapse bases \Footnote{With \emph{Haar random}, we refer to bases being drawn from a probability distribution that is uniform with respect to the Haar measure, i.e., invariant under unitary transformations. Of course, one can also use other probability distributions.} for each iteration step and lattice site \cite{Binder2015-92}. Note that in both cases, these measurements break the symmetry associated with the conservation of total magnetization $\hS^z_\tot=\sum_i\hS^z_i$ or total particle number $\hat{N}_\tot=\sum_i\hat{n}_i$, respectively, because the basis states are not symmetry eigenstates. Such METTS computations hence simulate the grand-canonical ensemble and symmetries can in general not be utilized. 

\subsection{Maximally mixing bases}
If we use a single collapse basis $\{|\vn\ket\}$ such as $\{\hS^z_i\}$ eigenstates for a spin system or $\{\hat{n}_i\}$ eigenstates for a system of bosons or fermions, there is no dynamics at all at infinite temperature ($\beta=0$). Starting from an arbitrary initial state $|\vn\ket$, transitions to all other basis states are impossible such that the METTS simulation is stuck in the state $|\vn\ket$.

Having the METTS dynamics at high temperatures in mind, we can minimize autocorrelation times by switching between collapse bases $\{|\vn\ket\}$ and $\{|\tilde{\vn}\ket\}$ for which the distribution of overlaps $|\bra\tilde{\vn}|\vn\ket|$ is as flat as possible. In other words, we want that all overlaps $|\bra\tilde{\vn}|\vn\ket|$ are as small as possible. This guarantees that, at least at high temperatures, transitions to many states are possible and of similar probability. For $\beta=0$ and a Hilbert space of dimension $\mc{D}$ \Footnote{$\mc{D}=d^N$ for the Hilbert space of an $N$-site system with single-site Hilbert space dimension $d$.}, an optimal combination of bases yields transition probabilities $p_{\vn\to\tilde{\vn}}=|\bra\tilde{\vn}|\vn\ket|^2=1/\mc{D}$ $\forall_{\vn,\tilde{\vn}}$ \Footnote{This is the case where the expansion coefficients of any state $|\vn\ket$ with respect to the second basis $\{|\tilde{\vn}\ket\}$ all have absolute value $1/\sqrt{\mc{D}}$.} and also $p^{(2)}_{\vn\to\vn'}:=\sum_{\tilde{\vn}}p_{\vn\to\tilde{\vn}}p_{\tilde{\vn}\to\vn'}=1/\mc{D}$ $\forall_{\vn,\vn'}$. More generally, we call a sequence of $K$ bases \emph{maximally mixing}, if $p^{(K)}_{\vn\to\vn'}=1/\mc{D}$ $\forall_{\vn,\vn'}$ at infinite temperature, where $p^{(K)}$ are the transition probabilities after $K$ steps, i.e., when having cycled once through all $K$ collapse bases.

One example of maximally mixing bases are the bases $\{|\vs\ket\}$ and $\{|\tilde{\vs}\ket\}$ of $\{\hS^z_i\}$ and $\{\hS^x_i\}$ eigenstates for a spin-$1/2$ system \cite{Stoudenmire2010-12}. Here, $|\sigma_i\ket\in\{|\ua\ket,|\da\ket\}$ and $|\tilde{\sigma}_i\ket=\frac{1}{\sqrt{2}}(|\ua\ket\pm|\da\ket)$. Unfortunately, this choice of bases is not applicable if we want to exploit the conservation of the total magnetization.

We may call a sequence of $K$ bases \emph{efficiently mixing} if, at infinite temperature, there are many nonzero $K$-step transition probabilities $p^{(K)}_{\vn\to\vn'}$ of comparable (small) amplitude for every $\vn$, and if the transitions grant ergodicity.

In the following, we give specific examples for maximally mixing or at least efficiently mixing collapse bases that are also applicable for symmetric METTS, i.e., when global quantum numbers are conserved.

\subsection{Symmetric collapse bases with efficient mixing}
Inspired by the discrete Fourier transform, one can construct maximally mixing bases for any $\mc{D}$-dimensional Hilbert space. With a first orthonormal basis $\{|x\ket\,|\,x=1,\dotsc,\mc{D}\}$, we can choose a second basis as
\begin{equation}\label{eq:fourier_basis}
	|\tilde{k}\ket := \frac{1}{\sqrt{\mc{D}}}\sum_{x=1}^{\mc{D}} e^{2\pi i kx/\mc{D}} |x\ket
	\quad\text{for $k=1,\dotsc,\mc{D}$,}
\end{equation}
which we call the \emph{Fourier basis} with respect to $\{|x\ket\}$.
As $|\bra\tilde{k}|x\ket|^2=1/\mc{D}$ $\forall{x,k}$ we have indeed $p^{(1)}_{x\to\tilde{k}}=1/\mc{D}$ and also $p^{(2)}_{x\to x'}=1/\mc{D}$ $\forall{x,x'}$ at infinite temperature. Note that we are free to reorder the states $|x\ket$. Hence, permutations can be used to construct different versions of the Fourier basis.

\begin{figure}[t!]
\includegraphics[width=0.89\columnwidth]{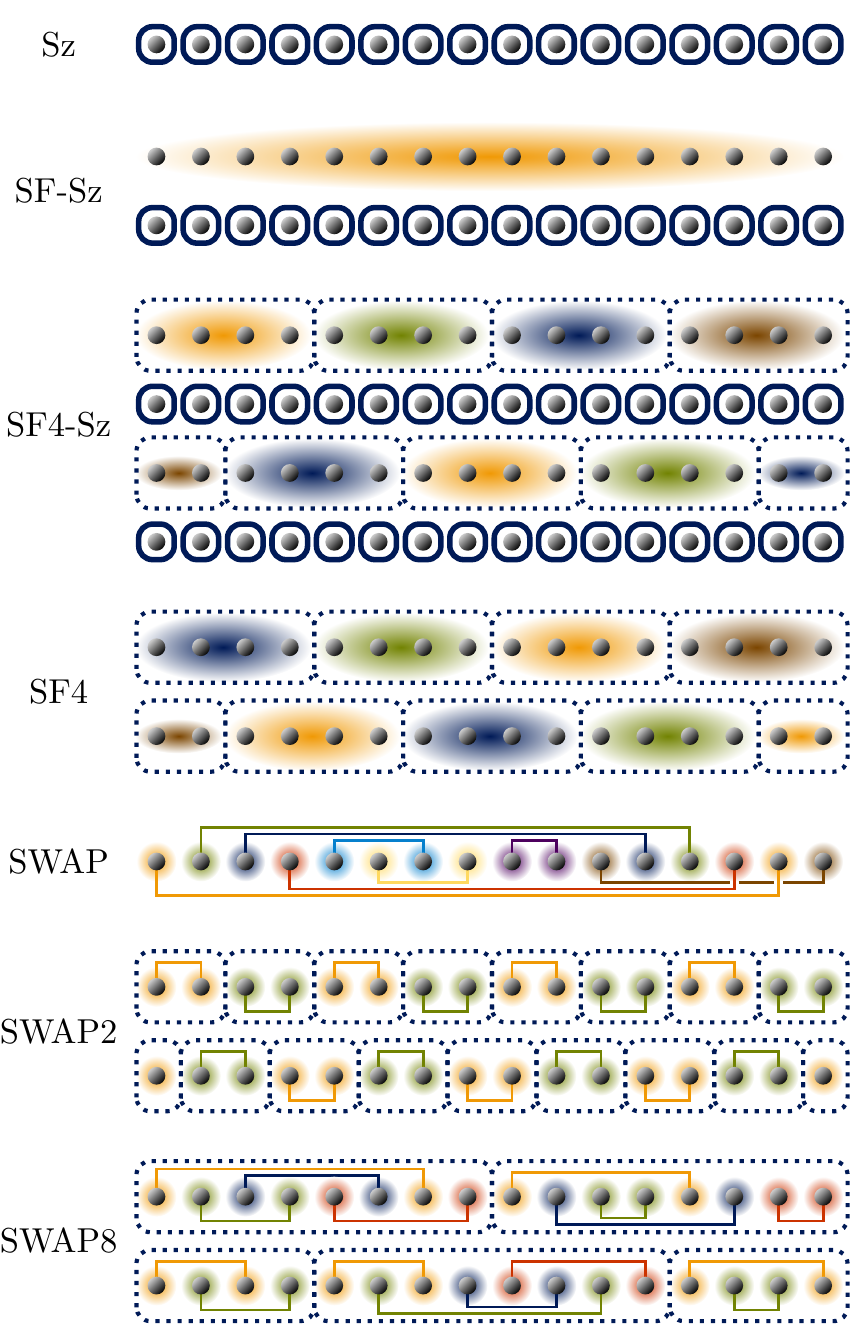}
\caption{Illustration of collapse bases. The top depicts the original collapse basis using the $\{\hS^z_i\}$ eigenbasis on each site (Sz). Below, we alternate between the Sz basis and its global symmetric Fourier transform \eqref{eq:fourier_basis} (SF-Sz). These are maximally mixing. In practice, we approximate the SF-Sz scheme by restricting the basis construction to blocks of $b$ sites, which are shifted by $b/2$ lattice sites for every second block collapse (SF$b$-Sz). Alternatively, a Haar-random symmetric basis can be chosen on these blocks (SR$b$-Sz). One can also omit the Sz-collapse in every second step and only use the Fourier basis, again shifting blocks in every second collapse (SF$b$). The swap collapse randomly partitions the lattice into pairs of sites, where, on each pair, the eigenstates of the swap operator \eqref{eq:swap_basis} are chosen. The overall collapse basis is the tensor product of these. Again, we approximate the ideal swap collapse by restricting the pairing to blocks of $b$ sites (SWAP$b$), where we shift by $b/2$ sites in every second collapse.}
\label{fig:collapse_drawing}
\end{figure}

In principle, we could use this approach to construct \emph{global} symmetric collapse bases that are maximally mixing. Let us discuss this using the example of a spin system.
Given an orthonormal basis $\{|\vs\ket\}$ of $\{\hS^z_i\}$ eigenstates for the magnetization $M$ subspace $\H_M$, we can identify the states $|\vs\ket$ with $|x\ket$, where $x=1,\dotsc,\dim\H_M$, and obtain their Fourier basis $\{|\tilde{k}\ket\}$ according to Eq.~\eqref{eq:fourier_basis}. These two bases of $\H_M$ are maximally mixing. We call the METTS scheme in which one alternates between collapses in this Fourier basis and the $\{\hS^z_i\}$ eigenbasis ``SF-Sz'' (\emph{Symmetric Fourier} w.r.t.\ $\hS^z_i$). It is illustrated in Fig.~\ref{fig:collapse_drawing}.

However, such a global Fourier transform is technically infeasible because the basis states $|\tilde{k}\ket$ are in general highly entangled. On average, their entanglement entropy is extensive, causing exponentially growing computation costs. To avoid this problem while retaining the efficient mixing property, we divide the lattice into blocks of $b$ lattice sites. For each of these blocks, we have an $\{\hS^z_i\}$ eigenbasis and can construct a Fourier basis for each symmetry sector of that block. This is scheme ``SF$b$-Sz'' in Fig.~\ref{fig:collapse_drawing}. For a spin-$1/2$ system with magnetization conservation, the $2^b$-dimensional Hilbert space of a $b$-site block is decomposed into $b+1$ symmetry sectors with magnetizations $-b/2, -b/2+1, \dots, +b/2$ and corresponding subspace dimensions $\tbinom{b}{0}, \tbinom{b}{1}, \dots, \tbinom{b}{b}$. For each of these subspaces, we construct collapse bases according to Eq.~\eqref{eq:fourier_basis} with $|x\ket$ referring in this case to the $\{\hS^z_i\}$ eigenstates of some fixed magnetization on the block (in some ordering). These bases are maximally mixing within each symmetry sector, but there are no transitions between different symmetry sectors of a block. Finally, in order to achieve ergodicity also at infinite temperature and to enhance the dynamics of fluctuations in general, we can shift blocks in every second block collapse by $b/2$ sites as illustrated in Fig.~\ref{fig:collapse_drawing}.

It is instructive to shortly discuss the nature of the resulting METTS dynamics for an example. The simplest case is a spin-$1/2$ system in the symmetry sector with a single up-spin, $M=-L/2+1$. First, consider infinite temperature, $\beta=0$. With global maximally mixing bases such as the symmetric Fourier $\hS_i^z$ bases (SF-Sz), every second collapse, the up-spin jumps with equal probability to any new site. In the corresponding block scheme SF$b$-Sz, the up-spin does a random walk (diffusive) whose average step size is proportional to the block size $b$. In contrast, for zero temperature ($\beta\to\infty$), every METTS is equal to the ground state in $\H_M$, for which, for Hamiltonians of interest, the up-spin is delocalized. So, at low temperatures, the change in position of the up-spin after each Sz collapse is not caused by the mixing property of the bases, but mainly by the delocalization due to the imaginary-time evolution.

Several variations of the above construction of efficiently mixing symmetric bases are conceivable. One is to use Haar-random collapse bases instead of the Fourier bases. So, instead of applying Eq.~\eqref{eq:fourier_basis} to obtain the second basis, one can draw a Haar-random basis for each symmetry sector of each $b$-site block. While such collapse bases, for the same block size $b$, perform very similarly at high temperatures, their efficiencies at finite temperatures may be quite different and will in general also depend on the system parameters. For the spin systems, we call the corresponding symmetric METTS schemes ``SR$b$-Sz''.

For both the SF$b$-Sz and SR$b$-Sz bases, where blocks are shifted in every second collapse, we noticed that in practice, the intermediate Sz collapses for every second sample are not really necessary. They actually result in somewhat slower convergence in our exemplary benchmark simulations. This can again be understood by considering the overlaps of the different basis states. We denote the schemes where Sz collapses are omitted by SF$b$ and SR$b$.

The symmetric Fourier and random block bases naturally have non-symmetric counterparts that can be applied in simulations of grand-canonical ensembles without symmetries. One simply omits the partitioning of the Hilbert space into symmetry sectors and uses Eq.~\eqref{eq:fourier_basis} or the Haar-random choice in the full $b$-site Hilbert space. Such non-symmetric Fourier block bases are always maximally mixing.

Let us briefly mention another variation that we pursued for the construction of efficiently mixing symmetric collapse bases. It is based on the swap operator that acts on a pair of lattice sites and swaps their quantum states. For two spins-$1/2$, it is
\begin{equation*}
	\SWAP = |\ua\ua\ket\bra\ua\ua| + |\ua\da\ket\bra\da\ua| + |\da\ua\ket\bra\ua\da| + |\da\da\ket \bra\da\da|.
\end{equation*}
Let $\SWAP_{i,j}$ act on sites $i$ and $j$ of the lattice. As $[\SWAP_{i,j}, \hS^z_\tot]=0$, we can use a symmetric eigenbasis of the swap operator,
\begin{equation}\label{eq:swap_basis}
	\{|\ua\ua\ket, \frac{1}{\sqrt{2}}\big(|\ua\da\ket + |\da\ua\ket\big), \frac{1}{\sqrt{2}}\big(|\ua\da\ket - |\da\ua\ket\big), |\da\da\ket\},
\end{equation}
as a collapse basis on pairs of sites. Ideally, we would randomly select arbitrary pairs of lattice sites $(i,j)$ and choose the swap eigenbasis \eqref{eq:swap_basis} on each of these pairs, forming the global collapse basis as their tensor product (``SWAP'' in Fig.~\ref{fig:collapse_drawing}). To avoid extensive entanglement, we again restrict the random pairing of sites to blocks of $b$ lattice sites (``SWAP$b$'' in Fig.~\ref{fig:collapse_drawing}).  As before, we also shift the blocks by $b/2$ lattice sites in every second collapse. For block size $b = 2$, the symmetric Fourier basis coincides with the swap basis.

\section{Factors influencing convergence speeds} \label{sec:factors}
Clearly, the model and the system parameters strongly affect the convergence of observables in the METTS sampling algorithm. For a given system, we can significantly influence the convergence properties by the choice of the collapse bases. This influence is mediated by two key mechanisms. First, the choice of the collapse bases $\{|\vn^{(\kappa)}\ket\}$ ($\kappa=1,\dotsc,K$) determines the METTS ensembles $\{|\phi^{(\kappa)}_\vn\ket\}$ from which we sample. If we could draw independent samples without any autocorrelations, the statistical error of an observable $\bra\hO\ket$ would solely depend on the distribution of its measurement values $\{\bra\phi^{(\kappa)}_\vn|\hO|\phi^{(\kappa)}_\vn\ket\}$ and their corresponding probabilities $P^{(\kappa)}_\vn$ in the METTS ensembles.
Second, the collapse bases influence the strength of autocorrelations between METTS samples.

To see the interplay of collapse bases and observables, consider, e.g., the operator $\hS^+_0\hS^-_3$ in a spin-$1/2$ system with spin-flip symmetry. At high temperatures ($\beta\to 0$), every METTS sample equals its corresponding basis state up to corrections of order $\beta$, i.e., $|\phi_\vn\ket = |\vn\ket + \mc{O}(\beta)$. Now, if we collapse into states $|\vn\ket$ that are eigenstates of $\hS^z_0$ and $\hS^z_3$, we have $\bra\vn|\hS^+_0\hS^-_3|\vn\ket = 0\ \forall\vn$ and, hence, $\bra\phi_\vn|\hS^+_0\hS^-_3|\phi_\vn\ket = \mc{O}(\beta)$. The distribution of measurement values in this METTS ensemble is peaked around the expectation value $\bra\hS^+_0\hS^-_3\ket=\mc{O}(\beta)$, leading to small statistical errors. If, however, we choose basis states $|\vec{m}\ket$ that are eigenstates of $\hS^x_0$ and $\hS^x_3$, we have $\bra\vec{m}|\hS^+_0\hS^-_3|\vec{m}\ket = \pm\frac{1}{4}\ \forall\vec{m}$, and $\bra\phi_{\vec{m}}|\hS^+_0\hS^-_3|\phi_{\vec{m}}\ket = \pm\frac{1}{4} + \mc{O}(\beta)$ with probabilities $\frac{1}{2}$, leading to large statistical errors. Note that if our observable of interest happened to be, say, $\hS^z_0$, the effect would be exactly reversed. Hence, there is no generally optimal choice of collapse bases as the answer also depends on the observables of interest. At low temperatures, the METTS become similar to the ground state of the system and, provided it is non-degenerate, the distribution of measurement values is strongly peaked around the ground state expectation value for any collapse basis.

\begin{figure}[t!]
 \includegraphics[width=\columnwidth]{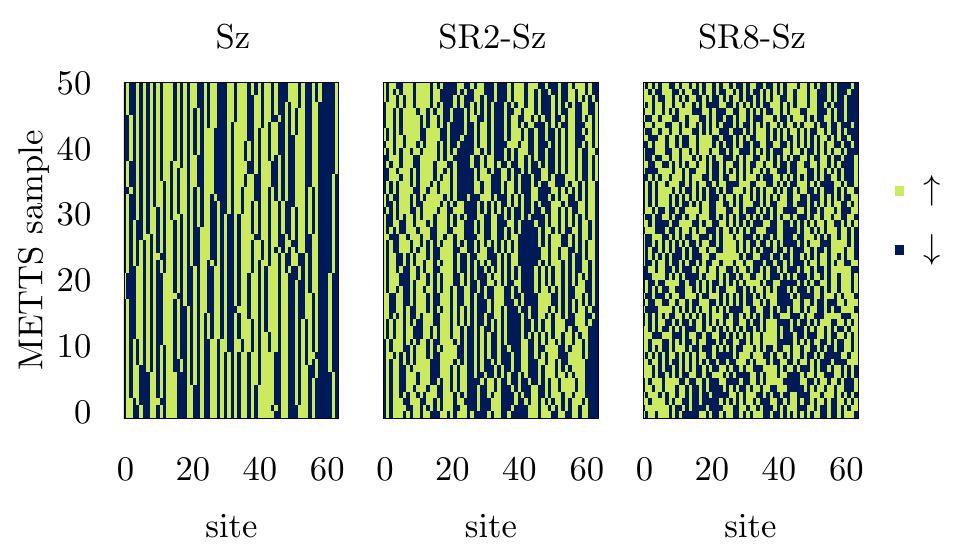}
 \caption{METTS sampling for the canonical ensemble of an isotropic spin-$1/2$ Heisenberg antiferromagnet at inverse temperature $\beta=1$, zero magnetization, and with system size $L=64$. The figure illustrates the effect of different symmetric collapse bases: Sz collapses only (left), and alternating Sz and symmetric Haar-random collapses (SR$b$-Sz) with block sizes $b=2$ (center) and $b=8$ (right). The panels show the $\{\hS^z_i\}$ eigenstates after Sz collapses. Thermalization of the Markov chains was ensured by discarding the first $1000$ samples.}
 \label{fig:diffusion}
\end{figure}

For the effect of autocorrelations, consider first the na\"ive way of simulating canonical ensembles by using the same symmetry eigenbasis $\{|\vn\ket\}$ for every collapse (e.g., ``Sz'' in Fig.~\ref{fig:collapse_drawing}). At infinite temperature, as discussed previously, the Markov chain cannot leave its initial state because $p_{\vn\to\vn'} = |\bra\vn'|\phi_\vn\ket|^2 = \delta_{\vn,\vn'}$. For small values of $\beta$, the probability that two subsequent samples are equal is still high because $|\bra\vn'|\phi_\vn\ket|^2=\delta_{\vn,\vn'}+\mc{O}(\beta)$. Thus, the collapse may occasionally induce a few transitions, but autocorrelations between subsequent samples remain high. One obtains slow diffusive dynamics in the Markov chain (left panel in Fig.~\ref{fig:diffusion}). This can be resolved by using a sequence of efficiently mixing bases. For a spin-$1/2$ system with magnetization conservation, we can, e.g., alternate between Sz collapses and the corresponding symmetric Fourier or Haar-random block bases. The larger we choose the blocks, the faster the Markov chain explores the state space due to reduced autocorrelations (center and right panel in Fig.~\ref{fig:diffusion}).

The influence of the system parameters becomes more important at lower temperatures. As the temperature is decreased, the METTS $|\phi_\vn\ket$ get closer and closer to the ground state. Typically, deviations from the ground state are localized if the system is gapped, and delocalized if the system is critical. For our block collapse bases, localized deviations then lead again to diffusive dynamics in the METTS and autocorrelations may considerably depend on specifics of the chosen bases. Delocalized deviations generally lead to short autocorrelation times.

\section{Results for XXZ chains in the canonical ensemble} \label{sec:CE-XXZ}
\begin{figure*}[t]
 \includegraphics[width=0.98\textwidth]{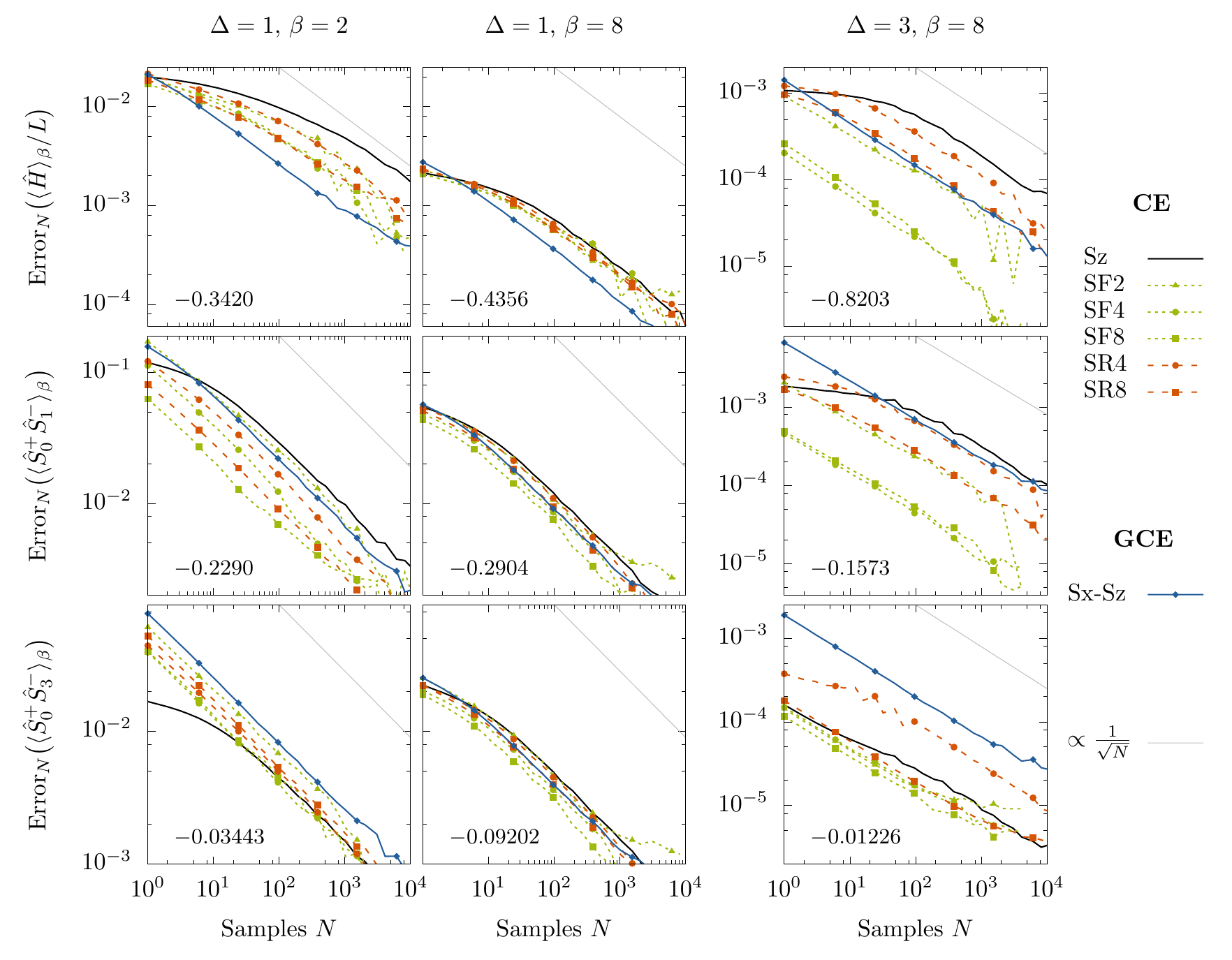}
 \caption{Convergence of METTS for different symmetry-conserving collapse bases in the CE. We study spin-$1/2$ XXZ chains with $L=64$ sites and zero magnetization at the isotropic point (left two columns, anisotropy parameter $\Delta=1$ in Eq.~\eqref{eq:H_XXZ}) and in the gapped phase ($\Delta=3$) for different inverse temperatures $\beta$. The considered observables are the energy per site $\bra\hH/L\ket$ (top), the correlator $\bra\hS^+_0\hS^-_1\ket$ (center), and the correlator $\bra\hS^+_0\hS^-_3\ket$ (bottom), where site $i=0$ refers to the center of the lattice. For the collapse bases, we compare the $\{\hS^z_i\}$ eigenbasis on each site (Sz), the symmetric Fourier (SF$b$), and the symmetric random (SR$b$) bases on blocks of $b$ sites. As a reference, we also show the convergence for alternating $\{\hS^x_i\}$ and $\{\hS^z_i\}$ eigenbases (Sx-Sz) in a simulation of the GCE. Numbers in the lower left corners of the panels state the quasi-exact values of the observables in the CE.}
 \label{fig:collapse_convergence}
\end{figure*}

In the following, we demonstrate the influence of the collapse bases on convergence speeds for spin-$1/2$ XXZ chains \cite{Bethe1931,Cloizeaux1966,Mikeska2004} with Hamiltonian
\begin{equation}\label{eq:H_XXZ}
	\hH = \sum_i \left(\hS^x_i \hS^x_{i+1} + \hS^y_i\hS^y_{i+1} + \Delta \hS^z_i \hS^z_{i+1} \right),
\end{equation}
with varying values of the anisotropy parameter $\Delta$. We focus on symmetric METTS computations for the canonical ensemble (CE) with zero magnetization and also compare the convergence speeds to those of non-symmetric simulations of the grand-canonical ensemble (GCE) with $\bra\hS^z_\tot\ket=0$. The convergence behavior for different temperatures, observables, and collapse bases is shown in Fig.~\ref{fig:collapse_convergence}. It demonstrates how our novel collapse bases can significantly improve the convergence in practice.

\begin{figure}[t!]
 \includegraphics[width=0.9\columnwidth]{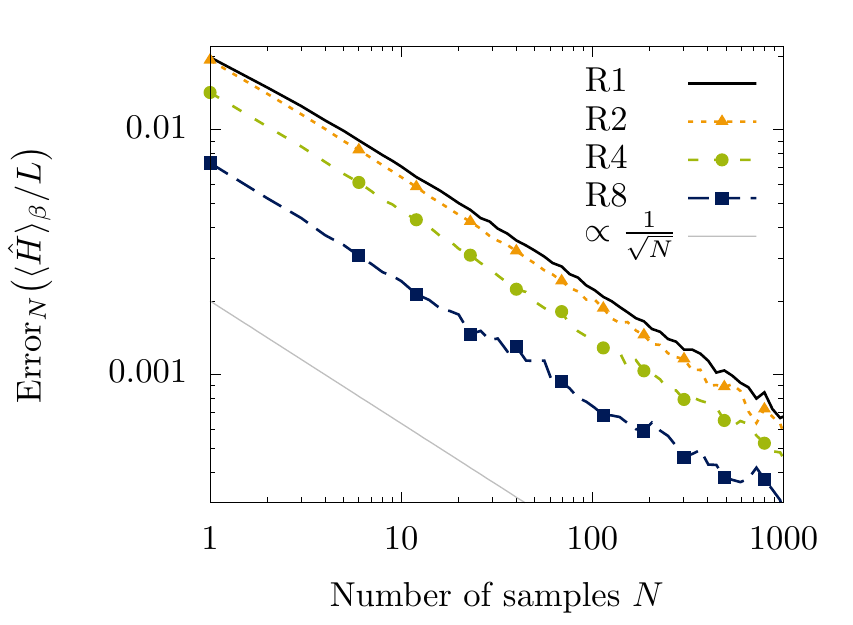}
 \caption{Influence of collapse bases on the convergence in the GCE without symmetries for the antiferromagnetic spin-$1/2$ Heisenberg chain ($\Delta=1$ in Eq.~\eqref{eq:H_XXZ}) at inverse temperature $\beta=1$. We show METTS errors for the energy per site $\bra\hH/L\ket$ as a function of the number of samples $N$. The curves compare non-symmetric Haar-random collapse bases (R$b$) on blocks of $b=1$, $2$, $4$, and $8$ sites.}
 \label{fig:gce_without_symmetries}
\end{figure}

The statistical error of the METTS sampling is quantified as follows. First, we apply the purification approach \cite{Verstraete2004-93}, which has been adapted to also describe canonical ensembles and utilize symmetries \cite{Nocera2016-93,Barthel2016-94}, to compute reference values of observables. For these simulations, we use the same system parameters, ensemble, and system size as for the corresponding METTS computation, and set a very low DMRG truncation threshold such that we can regard the obtained reference values as quasi-exact. For the error of $N$ METTS (Error$_{N}$ in Figs.~\ref{fig:collapse_convergence}, \ref{fig:gce_without_symmetries}, and \ref{fig:gce_convergence_ensembles}), we generate several sets of $N$ subsequent samples, and take the root mean square of the average (absolute) deviations of the observable from the quasi-exact reference data in each set \Footnote{In practice, when reference values are not available, one can apply standard techniques for the estimation of statistical errors in Monte Carlo simulations, e.g., jackknife resampling or bootstrapping, to probe the efficiency of different collapse bases.}. For the METTS simulations, the DMRG truncation weight $\epsilon$ and the Trotter time step $\Delta\tau$ were chosen such that the corresponding errors are negligible compared to the statistical error. See, e.g., Ref.~\cite{Binder2015-92} for details or Ref.~\cite{Schollwoeck2005} for a general review.

In Fig.~\ref{fig:collapse_convergence}, the first two columns show results for the isotropic antiferromagnet ($\Delta=1$ in Eq.~\eqref{eq:H_XXZ}) at inverse temperatures $\beta=2$ and $8$, and the third column shows results for $\Delta=3$ in the gapped N\'{e}el phase at inverse temperature $\beta=8$. The rows correspond to different observables: the energy per site $\bra\hH/L\ket$ and the correlators $\bra\hS^+_0\hS^-_1\ket$ and $\bra\hS^+_0\hS^-_3\ket$. Site $i=0$ is located at the center of the lattice, i.e., correlators are evaluated near the center to minimize finite-size effects \Footnote{Specifically, site 0 is the left of the two central sites for even system sizes $L$. We tested the influence of the exact position of site 0 on the correlators. It is negligible except for very high temperatures. Also, one could exploit spatial self-averaging \cite{Binder2015-92}, but we did not in this study.}. At high temperatures ($\beta=2$), using only the Sz collapse leads to strong autocorrelations that slow down the convergence for all three observables shown here. This problem can be alleviated by choosing the efficiently mixing symmetric Fourier or symmetric Haar-random collapse bases. In many cases, the symmetric Fourier bases work slightly better than the symmetric random bases. Errors reduce with increasing block sizes $b$. Larger block sizes typically also increase the computation costs per sample. For the correlator $\bra\hS^+_0\hS^-_3\ket$, the simple Sz collapse leads, for the chosen temperatures, to relatively small statistical errors. This can be attributed to the fact that the distribution of measurement values is peaked around the (small) expectation values $\bra\hS^+_0\hS^-_3\ket$, while it is broader for our novel block bases SF$b$ and SR$b$. Still, the latter reduce autocorrelations between samples: The curves corresponding to our new block bases roughly follow the $1/\sqrt{N}$ convergence, while the Sz-curves start rather flat, which is due to the autocorrelations. In the combination of both effects, the novel block bases typically outperform the simple Sz collapse, sometimes reducing errors by an order of magnitude.

For the critical system at low temperatures (center column in Fig.~\ref{fig:collapse_convergence}), the METTS errors are relatively independent of the chosen collapse bases. Our interpretation is that, in this case, the METTS are similar to the ground state with deviations from it being delocalized such that the particular choice of the collapse bases in the blocks is not as decisive.
For the gapped phase at low temperatures (right column in Fig.~\ref{fig:collapse_convergence}), however, the METTS errors vary significantly for the different bases. Our interpretation is that, in this case, deviations from the ground state are much more localized such that the distributions of measurement values and autocorrelations depend considerably on the basis choice.
For both values of $\Delta$, the Fourier bases (SF$b$) provide the best results.

To compare convergence speeds, Fig.~\ref{fig:collapse_convergence} also shows METTS errors for a simulation of the GCE using the non-symmetric METTS algorithm, where one alternates between the $\{\hS_i^z\}$ and $\{\hS_i^x\}$ collapse bases. The comparability is of course somewhat limited as the CE and GCE are not equivalent for our finite systems.
Overall, the statistical errors in the simulation of the CE with our novel symmetric collapse bases are comparable to and sometimes considerably smaller than those of the GCE simulation using the Sx-Sz bases. Hence, beyond cases where one specifically wants or needs to study the CE, simulating the CE with symmetries can be an efficient variant of the METTS algorithm when one is interested in the thermodynamic limit. This is possible because, as we learn from statistical mechanics, the different equilibrium ensembles such as the CE \eqref{eq:CE} with quantum number $Q$ and the GCE \eqref{eq:GCE}, with $\alpha$ tuned such that $\bra\hQ\ket^\gce_\alpha=Q$ are equivalent in the thermodynamic limit.

The symmetry-breaking versions of the Fourier and Haar-random bases can also be applied in simulations of the GCE without symmetries. In many cases, this improves the convergence of the sampling as illustrated for the antiferromagnetic spin-$1/2$ Heisenberg chain in Fig.~\ref{fig:gce_without_symmetries}. Here, we use Haar-random symmetry-breaking collapse bases (R$b$) on blocks of $b=1,2,4,$ and $8$ sites and compare the statistical errors of the thermal energy per site. The behavior observed here is typical: The statistical errors reduce as the block size is increased. Note, however, that larger block sizes also tend to increase the computation cost per sample.

\section{Symmetric simulation of grand-canonical ensembles} \label{sec:GCE}
Except for Ref.~\cite{Lacki2015-91}, METTS have so far been used exclusively to simulate grand-canonical ensembles \eqref{eq:GCE} \cite{White2009-102,Stoudenmire2010-12,Alvarez_2013,Bonnes_2014,Binder2015-92,Bruognolo2015-92}, alternating between different collapse bases that break the symmetry of the system to reduce autocorrelations. In this case, basis states are not eigenstates of the conserved quantities ($\hQ$) and it is hence in general not possible to exploit symmetries in the MPS representation \eqref{eq:MPS} of the METTS.

In the following, we discuss a modification of the algorithm to simulate grand-canonical ensembles under utilization of symmetries.

\subsection{The symmetric METTS algorithm}
\begin{figure*}[t]
\includegraphics[width=0.98\textwidth]{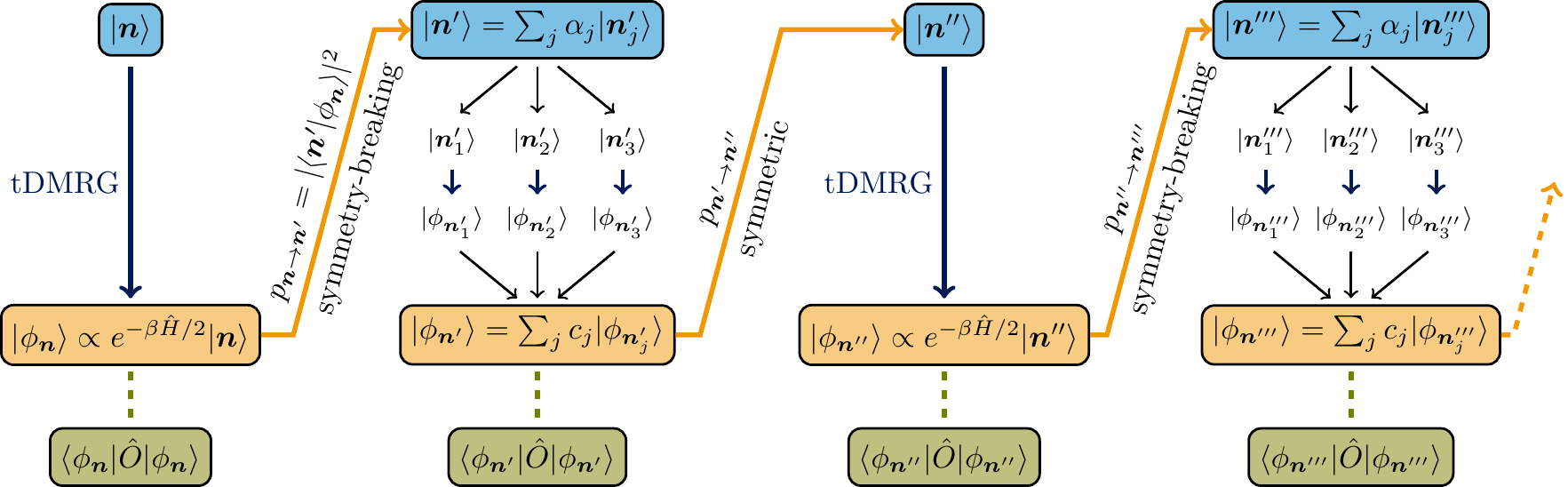}
\caption{Illustration of our symmetric METTS algorithm for the GCE using small sets of symmetry eigenstates. An initial symmetric basis state $|\vn\ket$ is evolved in imaginary time to obtain the METTS $|\phi_\vn\ket$. After the evaluation of observables, a suitable symmetry-breaking collapse results in a non-symmetric basis state $|\vn'\ket$. It is split into its symmetric components $\{|\vn_j'\ket\}$, which are separately evolved to obtain a small collection of symmetric states $\{|\phi_{\vn_j'}\ket\}$. Observables are evaluated with respect to this collection according to Eq.~\eqref{eq:superpos_observable_symmetric}, and a collapse using a symmetric basis yields a new symmetric basis state $|\vn''\ket$. We keep alternating between symmetry-breaking and symmetric collapse bases to explore different symmetry sectors according to their weights in the GCE.}
\label{fig:gce_scheme_ensembles}
\end{figure*}
As illustrated in Fig.~\ref{fig:gce_scheme_ensembles}, we employ small collections of symmetric METTS to exploit symmetries in the DMRG time evolution. Starting from an initial symmetric basis state $|\vn\ket$ with some quantum number $Q$, we use imaginary-time evolution to obtain the corresponding symmetric METTS sample $|\phi_\vn\ket$, where symmetries can be utilized in the tDMRG algorithm as described in Sec.~\ref{sec:symmetries}. After the evaluation of observables, we apply a collapse that breaks the symmetry in such a way that the resulting basis state $|\vn'\ket$ only has contributions from a small number of different symmetry sectors. We write the new basis state as a superposition of (normalized) symmetry eigenstates,
\begin{equation}\label{eq:symmComponents}
	|\vn'\ket = \sum_j \alpha_j|\vn'_j\ket,
\end{equation}
where $|\vn'_j\ket$ is the component with quantum number $Q_j$. Subsequently, the $|\vn'_j\ket$ are evolved in imaginary time separately, again exploiting symmetries.

The DMRG time evolution \cite{White2004,Daley2004} usually entails a Trotter decomposition. After each time step $s\Delta\tau\to (s+1)\Delta\tau$, the MPS should be renormalized to avoid numerical problems due to the exponential norm decay. So, after each time step, we multiply the evolved state $j$ by a factor $1/r^{(j)}_s$.
In order to keep the different renormalizations in the separate time evolutions of the states $|\vn'_j\ket$ consistent, we have to determine their relative weights $P_{\vn'_j} = \bra\vn'_j|e^{-\beta\hH}|\vn'_j\ket$. These can be obtained by multiplying the renormalization factors $r^{(j)}_s$,
\begin{equation}
	P_{\vn'_j} = \Big[\prod_s r^{(j)}_s\Big]^2 = \exp\Big[2\sum_s\log r^{(j)}_s\Big].
\end{equation}
As indicated, we do not actually multiply the $r^{(j)}_s$ but rather accumulate the sum of their logarithms because the $P_{\vn'_j}$ decay exponentially in the system size and inverse temperature $\beta$.
One should take care that the factors $r^{(j)}_s$ do not comprise the norm change due to the DMRG truncations of bond dimensions but only the norm change due to the application of evolution operators $e^{-\Delta\tau \hH}$. 

With the resulting normalized states $|\phi_{\vn'_j}\ket$, the normalized METTS sample is then given by
\begin{equation*}
	|\phi_{\vn'}\ket = \sum_j c_j |\phi_{\vn'_j}\ket, \quad \text{where} \quad 
	c_j \equiv \frac{\alpha_j\sqrt{P_{\vn'_j}}}{\sqrt{\sum_k |\alpha_k|^2 P_{\vn'_k}}},
\end{equation*}
and $|\phi_{\vn'_j}\ket = P_{\vn'_j}^{-1/2}\,e^{-\beta\hH/2}|\vn'_j\ket$. However, note that we can evaluate any observable $\hO$ without explicitly encoding the superposition $|\phi_{\vn'}\ket$ as an MPS according to $\bra\phi_{\vn'}|\hO|\phi_{\vn'}\ket = \sum_{kj} c_k^*c_j \bra\phi_{\vn'_k}|\hO|\phi_{\vn'_j}\ket$. For symmetry-conserving observables $[\hO,\hQ]=0$ we have $\bra\phi_{\vn'_k}|\hO|\phi_{\vn'_j}\ket = 0$ for $k\neq j$ as $Q_k\neq Q_j$ $\forall\ k\neq j$, and the expression reduces to the simple sum
\begin{equation}\label{eq:superpos_observable_symmetric}
	\bra\phi_{\vn'}|\hO|\phi_{\vn'}\ket = \sum_{j} |c_j|^2 \bra\phi_{\vn'_j}|\hO|\phi_{\vn'_j}\ket.
\end{equation}

Subsequently, the small collection of symmetric METTS is collectively collapsed to a new single symmetric basis state $|\vn''\ket$, using, e.g., any of the symmetric collapse bases described in Sec.~\ref{sec:bases}. The transition probabilities for this projective measurement can be obtained using $\hO=|\vn''\ket\bra\vn''|$ in Eq.~\eqref{eq:superpos_observable_symmetric}. We end up with a single state $|\vn''\ket$ that has one of the quantum numbers $\{Q_j\}$. This procedure is repeated, alternating between symmetry-breaking and symmetric collapse bases to explore the different symmetry sectors according to their weights in the GCE until the estimates of observables have reached the desired accuracy.

\subsection{Suitable symmetry-breaking bases}
As described, we suggest to alternate between symmetric and symmetry-breaking collapse bases, where elements of the latter should only have components from a few different symmetry sectors to allow for an efficient simulation.

For a spin-$1/2$ system with conservation of the total magnetization, a simple choice is based on the Sz collapse. For even iteration steps, we can use the $\{\hS^z_i\}$ eigenbasis. For odd iteration steps, we randomly select $n_x$ sites on which the $\hS^x_i$ eigenbasis is used as the collapse basis, while the $\hS^z_i$ eigenbasis is used on all other sites (``Sz/Sx''). A new set of $n_x$ sites is drawn for every symmetry-breaking collapse. It is straightforward to obtain the symmetric components, $\{|\vn'_j\ket\}$ in Eq.~\eqref{eq:symmComponents}, of the non-symmetric basis states $|\vn'\ket$ in MPS form. Note that with the described choice of bases, the parameter $n_x$ is directly related to the number of states in the small collection of symmetric METTS described above. Specifically, the states $|\vn'\ket$ have $n_x + 1$ components with different magnetizations that are separately evolved in imaginary time. Correspondingly, the total magnetizations of subsequent symmetric samples ($|\phi_\vn\ket$ and $|\phi_{\vn''}\ket$ in Fig.~\ref{fig:gce_scheme_ensembles}) differ by $\Delta M\in\{-n_x,\dots,n_x\}$.

In generalization of this symmetric Sz-Sz/Sx scheme, one can reduce autocorrelations, e.g., by starting from any of the efficiently mixing symmetric block bases discussed in Sec.~\ref{sec:bases} and, for odd iteration steps, modify the collapse basis for some randomly selected blocks to allow for changes of the conserved quantities, similar to the role of the $\hS^x_i$ eigenbasis above.

\subsection{Quantum number trajectories and convergence}
\begin{figure*}[t]
\includegraphics[width=\textwidth]{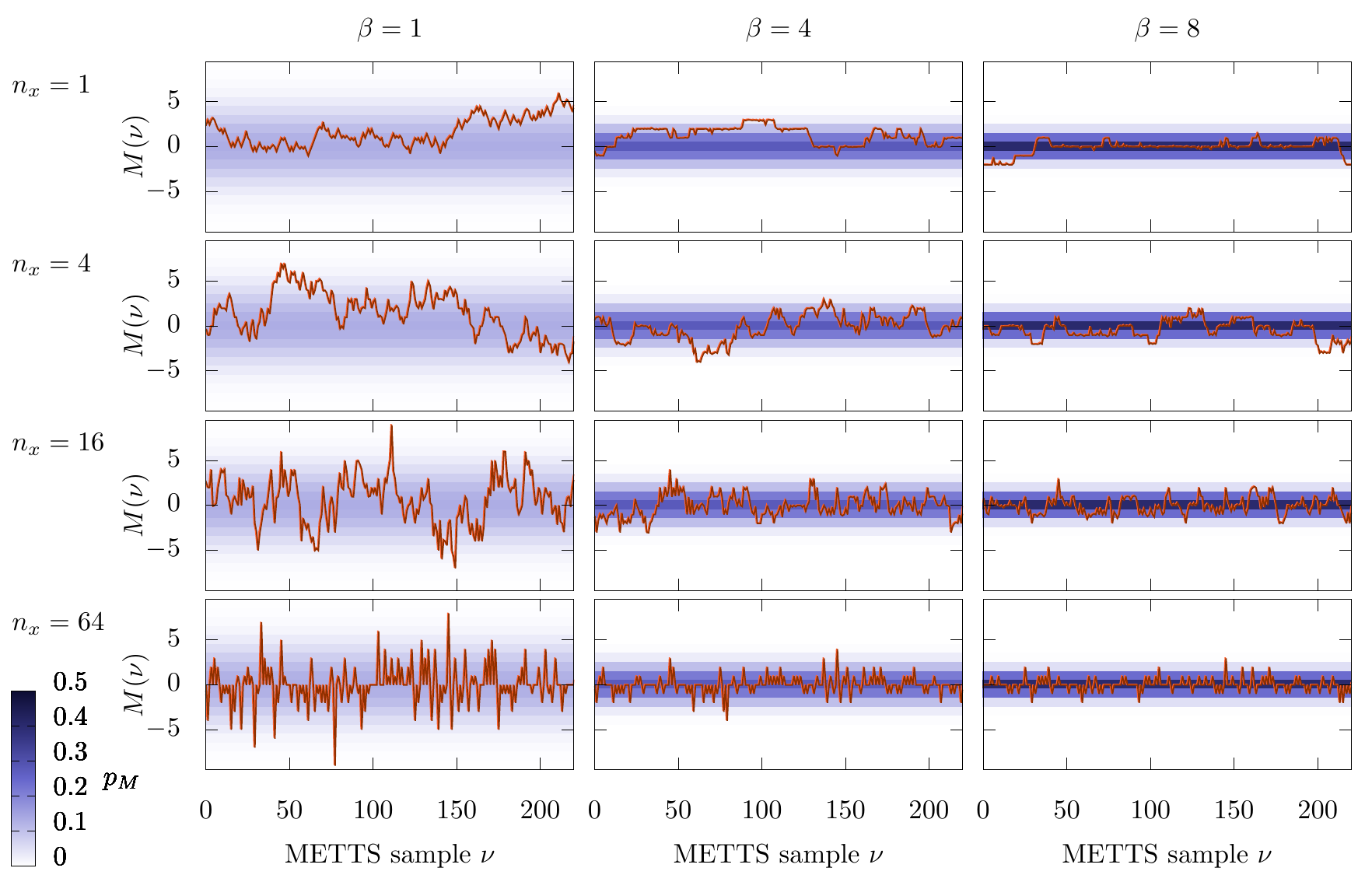}
 \caption{Simulations of grand-canonical ensembles for the antiferromagnetic spin-$1/2$ Heisenberg chain ($\Delta=1$ in Eq.~\eqref{eq:H_XXZ}) with $\bra\hS^z_\tot\ket=0$ and system size $L=64$. The density plots show the distribution $\{p_M\}$ of the total magnetization $M$, computed with MPDOs as described in Ref.~\cite{Barthel2016-94}. Lines show exemplary trajectories of $M(\nu)$ in Markov chains of symmetric METTS simulations using the SF4-Sz/Sx collapse bases as discussed in Sec.~\ref{sec:GCE}. They are characterized by $n_x$, the number of sites on which the symmetry-breaking $\hS^x_i$ eigenbasis is used. Note that for even iteration steps $\nu$, the state is an eigenstate of $\hS^z_\tot$ with quantum number $M(\nu)$ while, for odd $\nu$, we use the expectation value $M(\nu)=\bra\phi_{\vn^\nu}|\hS^z_\tot|\phi_{\vn^\nu}\ket$.}
 \label{fig:gce_paths_ensembles}
\end{figure*}
We test the algorithm for the isotropic spin-$1/2$ Heisenberg antiferromagnet, corresponding to $\Delta=1$ in Eq.~\eqref{eq:H_XXZ}. For collapses, we alternate between 4-site block symmetric Fourier bases (SF4 with blocks shifted by 2 sites in every second use), and nonsymmetric bases, where the $\hS^x_i$ eigenbasis is used for $n_x$ randomly selected sites and the $\hS^z_i$ eigenbasis for all other sites. We denote this combination of bases by ``SF4-Sz/Sx''.

Figure~\ref{fig:gce_paths_ensembles} shows trajectories of the total magnetization $M(\nu) = \bra\phi_{\vn^\nu}|\hS^z_{\tot}|\phi_{\vn^\nu}\ket$ occurring in the Markov chain and the probability distribution $\{p_M\}$ of the magnetization in the GCE, obtained by the MPDO approach that was introduced in Ref.~\cite{Barthel2016-94}. The rows in Fig.~\ref{fig:gce_paths_ensembles} show data for different values of $n_x$ in the symmetry-breaking collapse. The columns correspond to different inverse temperatures $\beta=1$, $4$, and $8$. Whenever we use the symmetric collapse basis (even steps $\nu$ in the Markov chain), the METTS sample is an $\hS^z_{\tot}$ eigenstate with magnetization $M(\nu)$. For odd $\nu$, $M(\nu)$ is the $\hS^z_{\tot}$ expectation value. Note that for $n_x=L$, we alternate between the $\{\hS_i^x\}$ eigenbasis and the SF4 bases (SF4-Sx collapse).

In all cases, the Markov chain appropriately explores the symmetry sectors according to their weights. At lower temperatures, the distribution is peaked around the center ($M=0$), while at higher temperatures, a large number of symmetry sectors contribute significant weight to the GCE. In the zero-temperature limit $\beta\to\infty$, where the system is in its ground state with quantum number $M=0$, the symmetry-breaking collapse forces the resulting superposition to have components with $M\neq0$. However, in this case, the imaginary-time evolution essentially projects onto the ground state, such that all contributions with $M\neq0$ are exponentially suppressed. The parameter $n_x$ determines the speed with which the Markov chain can explore the symmetry sectors. It limits the maximum change in magnetization that can be achieved in a single symmetry-breaking collapse.
\begin{figure*}[t]
\includegraphics[width=\textwidth]{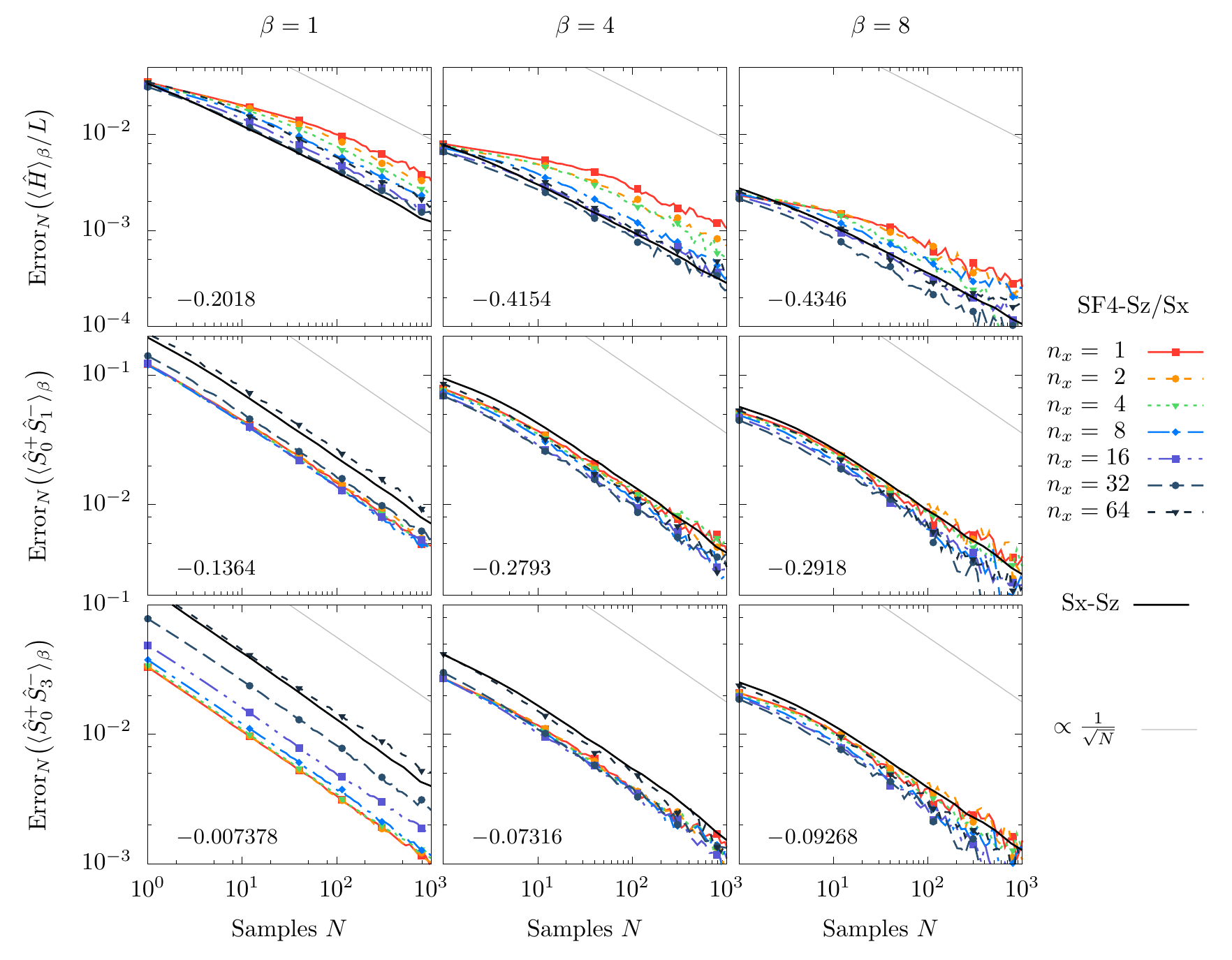}
\caption{Convergence of the symmetric METTS algorithm for grand-canonical ensembles, applied to the antiferromagnetic spin-$1/2$ Heisenberg chain with $L=64$ and $\bra\hS^z_\tot\ket=0$. We show the METTS error as a function of the number of samples $N$ at inverse temperatures $\beta=1$ (left), $4$ (center), and $16$ (right), for the energy density $\bra\hH/L\ket$ (top), the correlator $\bra\hS^+_0\hS^-_1\ket$ (center), and the correlator $\bra\hS^+_0\hS^-_3\ket$ (bottom). For collapses, we use the SF4-Sz/Sx bases as described in Sec.~\ref{sec:GCE} and compare their performance to the non-symmetric Sx-Sz collapse. Numbers in the lower left corners of the panels state the quasi-exact values of the observables.}
\label{fig:gce_convergence_ensembles}
\end{figure*}

Comparing with quasi-exact purification data \cite{Barthel2016-94}, Fig.~\ref{fig:gce_convergence_ensembles} shows the error of the algorithm as a function of the number of samples. The DMRG truncation weight $\epsilon$ and the Trotter step-size $\Delta\tau$ were again chosen such that the error is dominated by the statistical error. See Ref.~\cite{Binder2015-92} for a detailed discussion of the interplay of different error sources in the METTS algorithm. We consider the same observables as in Fig.~\ref{fig:collapse_convergence}, namely the energy per site $\bra\hH/L\ket$ and the correlators $\bra\hS^+_0\hS^-_1\ket$ and $\bra\hS^+_0\hS^-_3\ket$, all at inverse temperatures $\beta=1$, $4$, and $8$. The error of the energy expectation value decreases with increasing $n_x$ until $n_x=L/2$. For the nearest-neighbor correlator $\bra\hS^+_0\hS^-_1\ket$ the variations with $n_x$ are rather small. For the correlator $\bra\hS^+_0\hS^-_3\ket$ at the highest temperature ($\beta=1$), errors increase with increasing $n_x$, and, at the lower temperatures ($\beta=4,8$), the variations with respect to $n_x$ are again rather small. These properties can again be explained through the competition between autocorrelations and the spread in the distribution of measurement values.

In general, Fig.~\ref{fig:gce_convergence_ensembles} shows that also for rather small values of $n_x$, the error of our new symmetric METTS algorithm is comparable to the error of the original simulation of the GCE without the use of symmetries, where one alternates between $\{\hS^z_i\}$ and $\{\hS^x_i\}$ eigenbases on all sites (Sx-Sz collapse). Hence, the use of symmetries can make METTS simulations significantly more efficient.

\section{Conclusions and discussion} \label{sec:conclusions}
We have demonstrated how the METTS algorithm can be modified to allow for the utilization of symmetries. Conceptually, this is straightforward for the canonical ensemble -- one just needs to employ collapse bases that respect the symmetries of the system. In practice, it is important to cycle through a set of different collapse bases to reduce autocorrelation times. To this purpose, we have introduced and tested \emph{efficiently mixing} collapse bases such as Fourier bases and Haar-random bases which involve states that are entangled within blocks of several sites.
We have also introduced an efficient algorithm for the simulation of the grand-canonical ensemble under utilization of symmetries, using small collections of symmetric METTS.

Explicitly encoding symmetries in the MPS representation of the quantum states leads to a considerable speedup in the imaginary-time evolution and can hence make the sampling significantly more efficient. For both ensembles, we have discussed the effect of the collapse bases on the convergence of the METTS algorithm. Good bases grant short autocorrelation times in the Markov chain of METTS samples and a narrow distribution of measurement values for the observables of interest.

As demonstrated in this paper, understanding the role of the collapse bases and finding improved bases is a promising route to enhancing the efficiency of METTS simulations.

\end{document}